\documentclass[10pt,a4paper,twocolumn]{article}
\usepackage[latin1]{inputenc}
\usepackage[english]{babel}
\usepackage{amsmath}
\usepackage{authblk}
\usepackage{amsfonts}
\usepackage{amssymb}
\usepackage{natbib}
\usepackage{graphicx}
\title{Short and random: Modelling the effects of (proto-)neural elongations}
\author[a,b]{Oltman O. de Wiljes}
\author[c]{R.A.J. van Elburg} 
\author[a,b]{Fred A. Keijzer}
\affil[a]{Theoretical Philosophy, University of Groningen}
\affil[b]{Research School of Behavioural and Cognitive Neurosciences, University of Groningen}
\affil[c]{Institute of Artificial Intelligence, University of Groningen}
\date{}

\begin{document}
\twocolumn[
\begin{@twocolumnfalse}
\maketitle
\begin{abstract}
To understand how neurons and nervous systems first evolved, we need an account of the origins of neural elongations: Why did neural elongations (axons and dendrites) first originate, such that they could become the central component of both neurons and nervous systems? Two contrasting conceptual accounts provide different answers to this question. Braitenberg's vehicles provide the iconic illustration of the dominant input-output (IO) view. Here the basic role of neural elongations is to connect sensors to effectors, both situated at different positions within the body. For this function, neural elongations are thought of as comparatively long and specific connections, which require an articulated body involving substantial developmental processes to build. Internal coordination (IC) models stress a different function for early nervous systems. Here the coordination of activity across extended parts of a multicellular body is held central, in particular for the contractions of (muscle) tissue. An IC perspective allows the hypothesis that the earliest proto-neural elongations could have been functional even when they were initially simple short and random connections, as long as they enhanced the patterning of contractile activity across a multicellular surface. The present computational study provides a proof of concept that such short and random neural elongations can play this role. While an excitable epithelium can generate basic forms of patterning for small body-configurations, adding elongations allows such patterning to scale up to larger bodies. This result supports a new, more gradual evolutionary route towards the origins of the very first full neurons and nervous systems.
\end{abstract}
\\
\\
\end{@twocolumnfalse}
]

\section*{Introduction}
To understand how the very first neurons and nervous systems evolved, we need an account how each of neurons' most central characteristics came about: (a) their electrical signalling, (b) their synaptic connections and (c) their elongations (axons and dendrites). All three are central to nervous system functioning and each has evolved into a wide variety of forms and modes of operation within the huge group of animals now known as the neuralia~\cite[]{nielsen2008six}. From these three characteristics, graded and action potentials go back to unicellular organisms~\cite[]{naitoh1969ionic,greenspan2007introduction,Liebeskind2011}, and the same applies to macromolecular components of both the pre- and postsynaptic organization~\cite[]{burkhardt2015origin,Ryan_Grant_2009}. 

More difficulties remain with explaining how separate pre- and postsynaptic components came together to form synapses between separate cells, as well as how neural elongations first evolved. Both are tied to the multicellular morphology and functionality of proto- and early neuralia. To understand such early multicellular organizations, genomic and molecular evidence gives insufficient guidance~\cite[]{smith2014novel,nielsen2013life,hejnol2015acoelomorpha2}, while other evidence remains inconclusive. Fossils of neuralia go back to the beginning of the Cambrian, 542 Ma (Million years ago)~\cite[]{valentine2004origin} therefore synapses and neural elongations must have originated earlier. Body and trace fossils from the preceding Ediacaran (635 to 542 Ma) tend to be very different from modern animals and are difficult to interpret~\cite[]{budd2015early,fedonkin2007rise,Brasier_2009}. Molecular clock studies further suggest that the first neuralia evolved long before this period~\cite[]{cunningham2017origin,erwin2011cambrian}. Leaving no fossils, such proto- and early neuralia could have lived as small meiofauna with sizes up to 1 mm~\cite[]{wray2015molecular,erwin2015early}. It is also suggested that neurons and nervous systems evolved several times independently~\cite[]{moroz2009independent,moroz2014ctenophore}. At present, very little can be said with certainty about the first neuralia, either their form, size or how and when they lived. 

A systematic investigation and explication of potential evolutionary transitions from basic proto-neural configurations to neural ones will be beneficial here: this involves articulating possible trajectories that specify sequences of organismal organizations that span the transition from non-neural multicellular organizations to neural ones. Each step should consist of a functioning organism, while the consecutive steps from one organization to the next should be gradual, each one providing some improvement on the existing functionality~\cite[]{calcott2009lineage}. Investigating specific hypothetical transition trajectories will aid interpreting the limited empirical evidence and formulating more specific questions concerning this evidence. 

\begin{figure}
\begin{center}
\includegraphics[width=.8\linewidth]{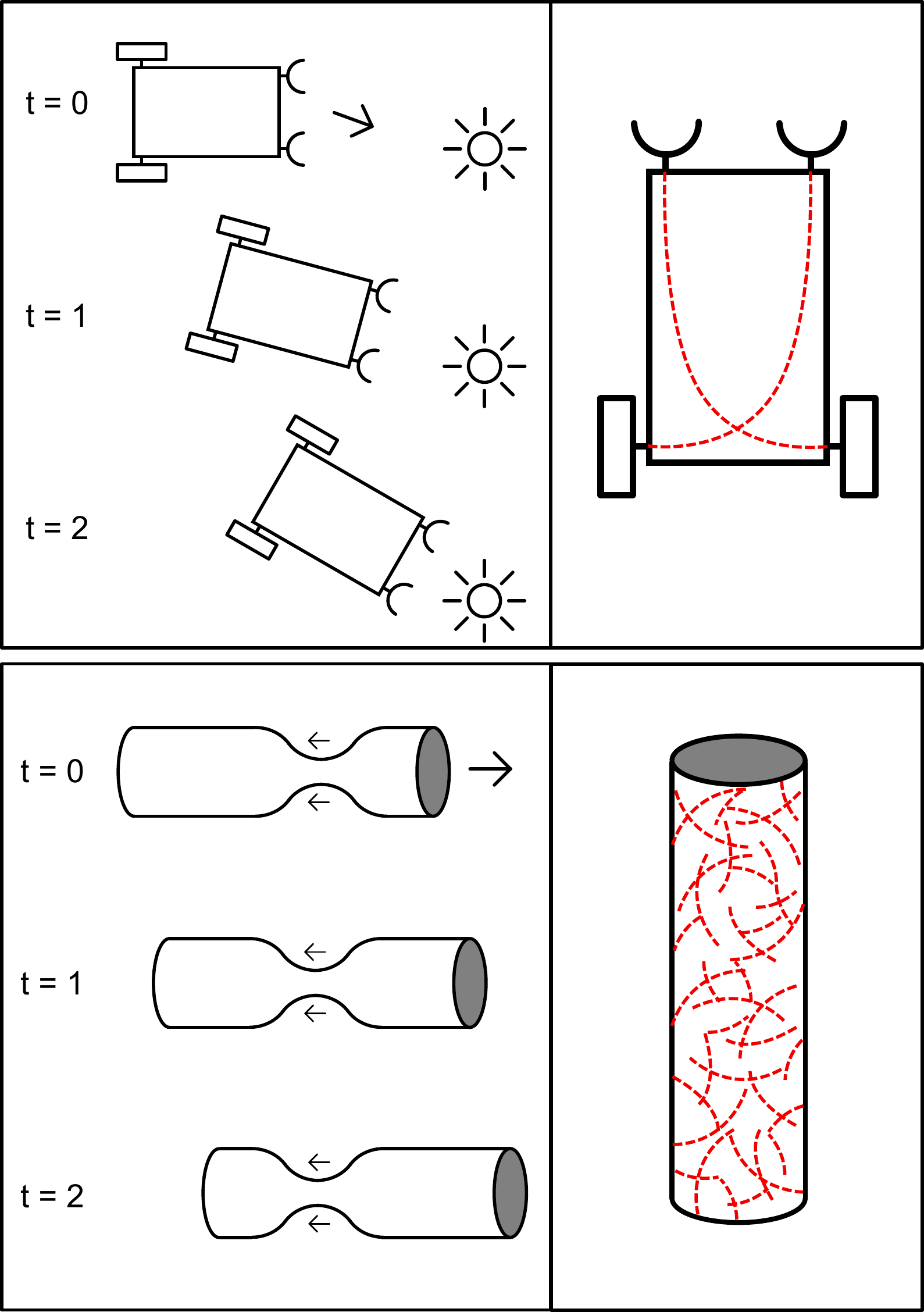}
\end{center}
\caption{A comparison between an input-output (IO) view as exemplified by a Braitenberg vehicle, top, and an internal coordination (IC) view, represented by an excitable (myo)epithelium with short and random connections, bottom. Behaviour of the respective systems is shown on the left; the wiring on the right.}
\label{fig:brait}
\end{figure}

A well-known example of such an idealized trajectory for nervous systems is provided by Braitenberg~\cite[]{braitenberg_1984}. He formulated a sequence of configurations starting with a single neural connection between a sensor and an effector to which more connections could be added, eventually leading to increasingly complex neural circuits. Figure~\ref{fig:brait}, for example, sketches a configuration with two connections. Braitenberg's proposed trajectory is based on an Input-Output (IO) view on (early) neural evolution~\cite[]{jekely2015option}. IO views stress the functioning of neurons---and whole nervous systems---as connections between sensors and effectors. Initially these connections may have been simple and direct, but over evolutionary time they have become increasingly complex neural circuits governing behaviour. Neural elongations function here as specific and often long-distance connections between specific loci within an organism (sensors, effectors, or other neurons). While IO views seem well-suited for modern nervous systems, they start with organizationally and developmentally complex bodily organizations, which raises doubts about their suitability as a primitive condition (but see~\cite{Jekely2011} for a possible approach). In addition, an IO view does not readily fit the most primitive extant nervous systems consisting of surface-distributed nerve nets such as those found in cnidarians and acoelomorpha~\cite[]{hejnol2015acoelomorpha2}.

In contrast, we focus here on an alternative Internal Coordination (IC) view~\cite[]{jekely2015option}, which stresses the need to acquire multicellular (bodily) coordination as an initial key task for early---and modern---nervous systems. Coordinating contractile (muscle) tissue for motility and reversible changes in body-shape is a central example here~\cite[]{pantin1956origin,keijzer2013nervous,keijzer2017animal} that imposes different functional demands on early nervous systems and neural elongations. Rather than focusing on neural elongations as a way to provide specifically targeted connections, an IC view opens up the possibility of acquiring neural elongations in a more gradual way.

We performed a modelling study to test the IC idea that simple---short and randomly connected---neural elongations can have played a significant behavioural role for proto-neuralia that we take here to be of limited (meiofaunal) size consisting of a few hundreds to thousands of cells. The starting point for this study consists of an excitable and contractile epithelium (a myoepithelium) that can provide primitive contractile motility~\cite[]{mackie1970neuroid}. Myoepithelia with electrical connections between the constituting cells exist in extant animals~\cite[]{mackie1968epithelial}. We hypothesize that proto-neuralia could have possessed myoepithelia with similar electrical connections or with chemical signaling to conduct electrical activity from one cell to the next, either by juxtacrine signaling or basic chemical synapses~\cite[]{keijzer2013nervous,wiljes2015modeling}. Importantly, this configuration could have provided a scaffold to bring separate pre- and post-synaptic elements in adjacent cells together as a full synapse. However, here we only focus on the impact of neural elongations on an excitable epithelium where electrical activity is transmitted from cell to cell (see figure~\ref{fig:brait}). 

The model is kept as abstract as possible: a worm-like body-tube consisting of a single tubular sheet of cells, representing an excitable epithelium; electrical excitability generic enough to represent either a calcium-based mechanism or a sodium/potassium based one, and abstract excitatory synaptic transmission to represent either electrical or chemical transmission. The model does not include external senses, proprioception, or pacemakers. A Poisson process randomly initiates activity in single cells. We consider traveling rings of activity across the body-tube that could enable a basic form of peristalsis to be relevant activity (Figure~\ref{fig:brait}).

\begin{figure*}
\begin{center}
\includegraphics[width=.70\linewidth]{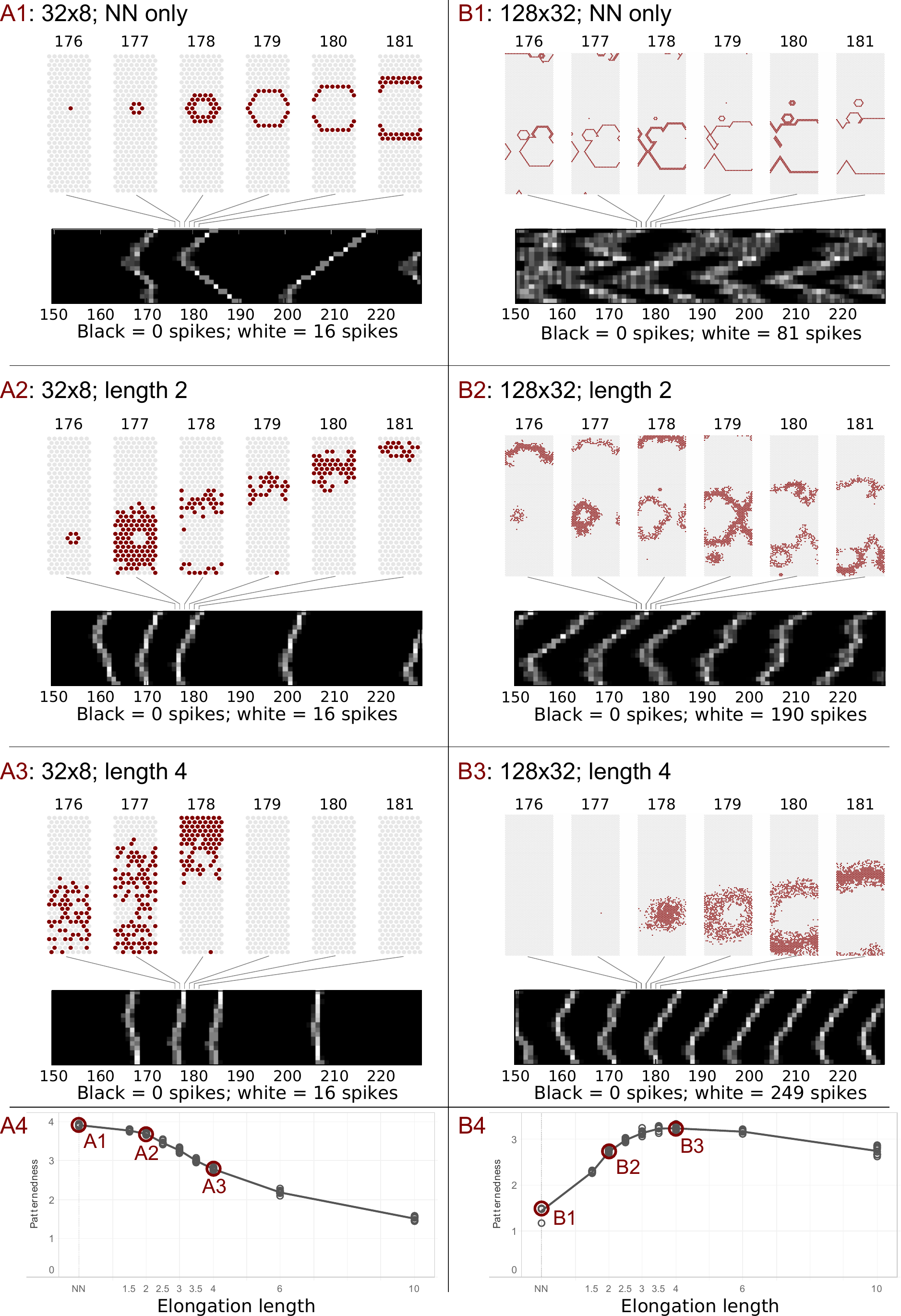}
\end{center}
\caption{Panes A1-3 and B1-3 each represent individual cases of sequences of activity across the body-tube for six different experiments (above), together with condensed representations spanning longer sequences (below). Individual time frames are identified by frame numbers, both above and below. Panes A1-3 show a small (32 by 8) body-tube, panes B1-3 a large one (128 by 32). Panes A1 and B1 represent nearest-neighbour connections only; A2 and B2 have in addition elongations of length 2; A3 and B3, have elongations of length 4. In all experiments shown, 50\% of the cells have elongations. The graphs in panes A4 and B4 represent the average patternedness for the conditions shown above them (marked in the graph) together with the results for five more elongation lengths. The points scattered around the graph represent the 10 individual model iterations which were performed for each condition. }
\label{fig:stripeComparison}
\end{figure*}

Earlier modelling work showed that an excitable epithelium provided a way to initiate ring-shaped spatiotemporal patterns of excitation running along the long axis of the body-tube ~\cite[]{wiljes2015modeling}. However, the presence of such patterns at the whole body scale depended on the form of the body-tube---a large length to width ratio was required---as well as the size of the body: ring-shaped activity only occurred for smaller tubes and was lost in larger ones. These patterns are an emergent feature resulting from the topology of the system, its limited size in terms of cells, random activations, and of the refractionary nature of the cells. For larger body-tubes, signalling to adjacent cells provides only travelling wave-fronts that remain thin and insignificant on the scale of the entire animal, while they do not travel fast enough to entrain excitation at the level of the whole body. 

With this setting, we investigated whether neural elongations could provide a mechanism to overcome the limitations of nearest neighbour signalling. We hypothesized that the patterns of activation across a body-tube would change when neural elongations were added. To keep the change as generic and basic as possible, we focused on the effects of elongations that are relatively short---that is, connecting cells that have only a few intermediate cells between them---while the connections are made randomly. Such short and random connectivity is undemanding in terms of body morphology, cell differentiation, and developmental patterning as it does not require preset destinations for these connections. Short and random elongations can therefore provide a small and plausible step in an evolutionary trajectory towards very primitive nervous systems. 

To test the validity of this idea we investigated a number of variations of randomly connected configurations: we varied the fraction of cells with elongations, the length of the elongations, and the size of the body-tube. The presence of ring-shaped patterns of excitation travelling along the length of the tube was used as a biologically plausible form of coordinated activity that also could be measured both qualitatively and quantitatively under various modelling conditions.

\section*{Results}
Central examples of our main results are presented in figure~\ref{fig:stripeComparison}. The panes on the left feature small systems measuring 32 cells in length and 8 in circumference. Those on the right represent large systems of length 128 and circumference 32. All systems have nearest neighbour (NN) signalling. Additionally, A2 and B2 have elongations of length 2, and A3 and B3 of length 4. These lengths refer to each cell's elongation length, implying that cells up to maximally 4 and 8 cells apart can become connected. Panes A1-3 and B1-3 each provide a detailed picture of the randomly initiated electrical activity across the cut open body-tube during six time steps of three milliseconds. Below is a condensed representation of this same activity across 100 time steps, providing a temporally extended overview of this activity by compressing all activity at a time-step into a line of 16 segments, scaled to the size of the body-tube (see figure~\ref{fig:stripeExplanation}). As white reflects high levels of activity, travelling waves are shown as diagonal lines, unpatterned activity as smudges, and synchronous activation of all cells as a vertical line. Together, these panes give an indication of both the details and the more abstracted differences between the patterning resulting from the various conditions.

In addition to these qualitative results, panes A4 and B4 show graphs for the same small and large body-tubes, representing more extensive parameter scans, involving the NN condition and eight different elongation lengths (on the x-axis). The y-axis shows the measure of patternedness calculated as outlined in the \ref{ch:PatQuant} section. The line represents the average patternedness for the given condition and the points the individual model runs. The conditions represented in detail in A1-3 and B1-3 are marked as such within the two graphs.

As previously found~\cite[]{wiljes2015modeling}, small systems without added elongations exhibit ring-shaped patterns (A1), while these patterns are lost when the system is larger (B1). As hypothesized, when neural elongations are added to such large systems, ring-shaped patterns return (B2-3). However, such elongations are detrimental for smaller systems (A2-3), where they have a negative effect on patternedness (A4). This can be seen both by inspecting the examples and the quantitative results presented in A4 and B4.

The comparison between A4 and B4 shows the main effect of elongations on different topologies: For the small systems, elongations lower the patternedness (A4). For the large systems, elongations increase the patternedness, though returns do diminish for the longer elongations (B4). 

\begin{figure}
\begin{center}
\includegraphics[width=\linewidth]{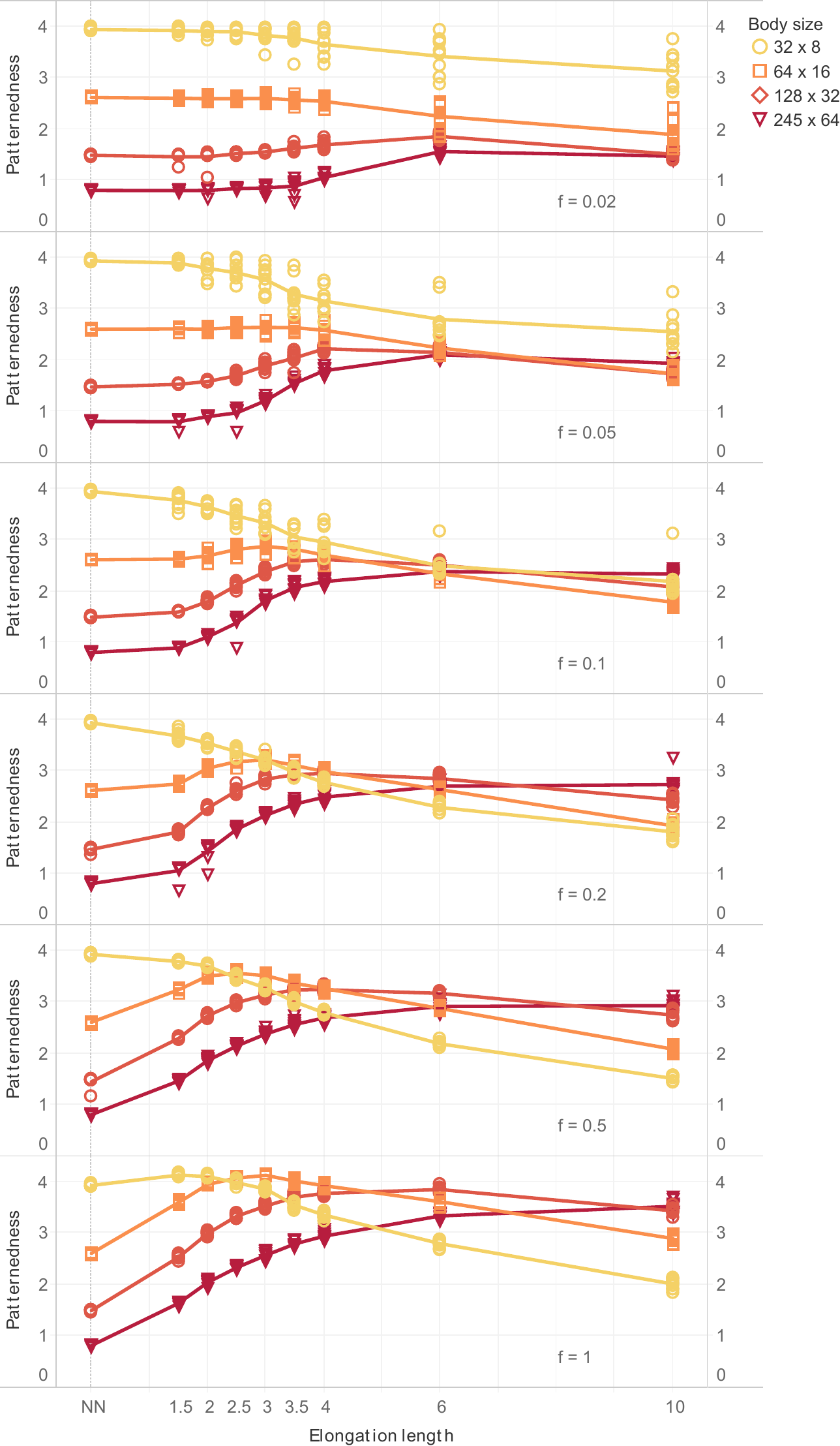}
\end{center}
\caption{Comparison of the effect on patternedness of different elongation lengths and body sizes for six elongation fractions, $f = .02$ through $f = 1.0$. Line color and point shape represents different body sizes. Patternedness is shown on the each individual y-axis, the different elongation lengths are on the x-axis and individual plots represent the various values of $f$. The points scattered around the graph line indicate the 10 individual model iterations which were performed for each condition and the line itself represents the average.}
\label{fig:errorBar_proportions}
\end{figure}

The results discussed so far are based on systems where the probability of each cell having an elongation is .5 (i.e. the chance that any given cell has an elongation on top of nearest-neighbour connectivity is 50\%) while we focused on two sizes of the body-tube. To investigate the effect of various elongation fractions (hereafter referred to as $f$) on patternedness, we also compared elongation fractions $f$ across various body sizes and elongation lengths as described above. We also did these experiments for two extra body sizes: 64 by 16 and 256 by 64. The results are presented in figure~\ref{fig:errorBar_proportions}, which is similar to panes A4 and B4 of figure~\ref{fig:stripeComparison} (included, respectively, as the lines marked with circles and diamonds in the pane marked `$f = 0.5$') but with additional conditions regarding $f$ as well as the additional body sizes. 

Figure~\ref{fig:errorBar_proportions} shows the effect of $f$ on patternedness. From top to bottom, $f$ increases. for larger systems (darker red lines, marked with diamonds and triangles), that having a higher fraction of cells with elongations improves patternedness. For the smallest system (yellow circles) elongations appear to be detrimental to patternedness. For the second-smallest system (orange squares), higher fractions (lower graphs) have a beneficial effect on patternedness for all but the longest two. However, low fractions (top graphs) show no such positive effect on patternedness---yet on the other hand, the low fractions also harm less for the longer elongation lengths.

Interestingly, for the larger body sizes, there appear to be diminishing returns to higher fractions. For the largest system (darkest line), patternedness for $f = 0.5$ is only marginally better than $f = 0.2$ for all elongation lengths. 

\section*{Discussion}

Our model represents an IC approach to early nervous system evolution (see figure~\ref{fig:brait}). Our aim was to investigate internal coherence scenarios that involve tissue configurations that are intermediate between non-neural and neural ones. In particular, we asked whether the presence of neural elongations providing random connections over very short distances could have functioned to enhance internal coordination for comparatively small meiofaunal proto-neuralia. Our results show how such elongations can indeed enhance coordinated activity across very basic multicellular configurations. We discuss four specific implications of the modelling results.

First, short and randomly directed neural elongations allow patterned activity within larger multicellular organisms. Direct connections between adjacent cells within an excitable epithelium can provide a way to initiate and maintain coordinated patterns of excitability across a body surface. Such patterns could have enabled organized contraction. However, while this mechanism works for small body-sizes, such patterning deteriorates for larger bodies~\cite[]{wiljes2015modeling}. The model presented here shows that short and random neural elongations can support similar forms of patterning for increasingly larger multicellular organisms. Thus such very primitive forms of neural elongations provide a mechanism for adapting patterned activity to changes in body-size. 

Second, neural elongations provide a way to scale the activity patterns themselves with respect to the size of the organism. An epithelial configuration only allows the spread of activation to adjacent cells, which limits the width of the patterns in the travelling direction to one or two cells (see Pane A1 and B1 of figure~\ref{fig:stripeComparison}). With elongations, the travelling patterns of activity can extend across more cells at every time-step. This allows the patterns themselves to become wider and to scale up with larger body sizes (see Pane B2 and B3).

Third, for larger systems (See Pane B4 of figure~\ref{fig:stripeComparison}), patternedness tends to correlate positively with the length of the elongations, although even short elongations already provide an improvement of patterning. Thus, while even short elongations can be beneficial, lengthening them over evolutionary time provides a gradual path to further improve such patterning capacities.

Fourth, while the influence of elongations on patterning tends to be more prominent when a larger fraction of cells have them, even a small fraction of cells with elongations can have significant effects depending on the size of the organism (see the top graph of figure~\ref{fig:errorBar_proportions}). Again, this provides an evolutionary path for a gradual improvement of the system.

Together these results provide a proof of concept that an organismal configuration relying on patterned activity across an excitable epithelium can use very basic neural elongations to maintain and improve patterning capacity for larger body-sizes. These changes can occur in small incremental steps allowing for a gradual evolutionary route towards increasingly complex neural elongations. 

These findings have a broad conceptual relevance for understanding the very early evolution of both neurons and nervous systems. Rather than assuming that neurons came first and through assembly constituted the first nervous systems, here the sequence is reversed. With the IC view developed here, `nervous system functioning' can be produced without full neurons by epithelia acting as a `proto-nervous system'. The latter can provide a scaffold for gradually evolving full neurons and nervous systems. In this way, tissue configurations spanning the gap between non-neural and neural tissues become conceivable. The model presented here shows that this speculative idea is indeed consistent and opens up new avenues for looking at the early evolution of nervous systems.

In comparison to the standard IO view that assumes the beneficial presence of long and specifically targeted neural connections, the results from this IC-based model provides new evolutionary scenarios that bring the neuron's three main features---electrical signalling, synapses, and elongations---together in a gradual and plausible way.

\section*{Materials and Methods}

\subsection*{Computational Model}

We reimplemented the model of \cite{wiljes2015modeling} using the brian2 package in Python~\cite[]{goodman2008brian}. All experiments use the following model:
\begin{itemize}
\item A 2D sheet of cells placed equidistantly in a triangular grid;
\item The sheet is rolled into a cylinder to create a tube;
\item Each cell is modelled by an integrate-and-fire model with a refractory period;
\item Each cell has superthreshold connections to its direct neighbors;
\item Each cell has a superthreshold per-cell Poisson process driver. In accord with \cite{wiljes2015modeling} the rate of Poisson process is set at 0.1 Hz; 
\item We used a worm-like topology. This is biologically plausible and functionally sufficient. The length to circumference ratio of the folded tube was fixed at 4:1. While we varied overall size leading to length by circumference combinations, respectively, of 32 by 8, 64 by 16, 128 by 32, and 256 by 64;
\item A fraction of the cells were given straight elongations in random directions which provide them with connectivity to each cell visited by the elongation; details can be found in supplement~\ref{sec:mikado}. We performed a parameter scan over various elongation lengths: 1.5, 2, 2.5, 3, 3.5, 4, 6, and 10 grid spacings. Additionally, we included a condition without any elongations, but as in all cases above keeping nearest neighbour connectivity;
\item Experiments have varying fractions of elongated cells. We performed a parameter scan over various elongated cell fractions. For fractions $< 1$, cells to be elongated were picked randomly. The fractions used are .02, .05, .1, .2, .5, and 1.
\end{itemize}

\subsection*{Pattern Quantification}
\label{ch:PatQuant}
The system as described above produces spatiotemporal patterns. Previous work shows that given suitable noise rates, nearest-neighbour connectivity, and activation with a refractory period, travelling ring-shaped activity patterns appear. To quantify the degree and relevance of ring-shaped activity on the surface of the animal we developed a pattern detection method. We detect patternedness 
by comparing the activity in ring-shaped segments with average activity over all segments. The activity in a segment is simply the number of cells active in a given ring-shaped segment in a given time frame. 

\begin{figure}
\begin{center}
\includegraphics[width=.6\linewidth]{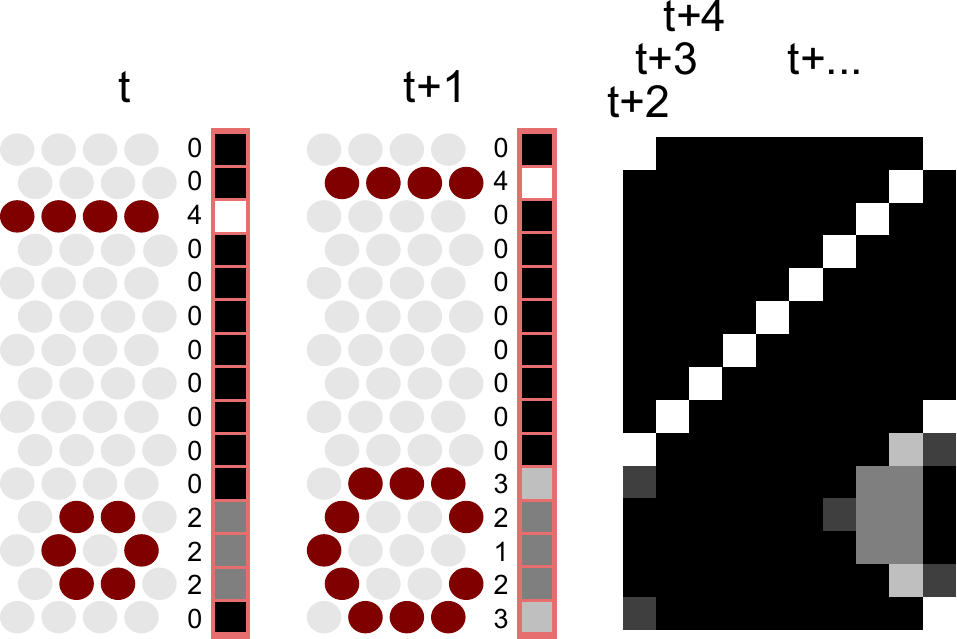}
\end{center}
\caption{This graphic shows how a condensed activity representation is calculated. Time steps t and t+1 show the process in detail. For each time-step, the body-tube---cut open in grey on the left---is divided into 16 ring-shaped segments (in this case, all 1 cell wide) for each of which the number of active cells---in red---is tallied. This numerical score is translated to greyscale values ranging from white (maximum) to black (minimum) providing a single vector for each time step.}
\label{fig:stripeExplanation}
\end{figure}

We summarize this activity for a given run as follows: 
\begin{itemize}
\item Divide the tube in $n$ segments. A cell $i$ is in segment $s$ if $(s-1) / n \cdot l_{tube} \leq x_{i} < s / n$.
\item A spike for a cell $i$ in segment $s$ takes place at a time $t$. The simulation is divided into time bins $[j\Delta t,(j + 1)\Delta t)$.
\item For each segment and time bin we can now count how many spikes take place: the matrix $C_{s,j}$ indicates the number of spikes in segment $s$ during time bin $j$.
\end{itemize}

We set $\Delta t$ to $3 ms$, larger than the average transmission delay between cells ($2 ms$) but shorter than the refractory period ($20 ms$). Within a time bin transmission between connected cells can occur but a single cell cannot fire twice. Empty and incomplete intervals at the start and the end of each run are discarded. The number of segments ($n$) is set to 16 for all systems in order to be able to compare measurements between the various body dimensions.

The ratio $T_{end} / \Delta t$ determines the number of intervals included in the analysis.

Visualizing $C_{s,j}$ as shown in figure \ref{fig:stripeExplanation} results in the condensed activity sequence plots found on the lower sides of panes A1-3 and B1-3 in figure~\ref{fig:stripeComparison}. Ring-shaped activity travelling down the tube shows up in this format as diagonal stripes, since a large amount of activation in a single ring-shaped segment travelling along the length of the animal and showing up in the next segment in a later time interval is a diagonal move.

To quantify the deviation in activity we performed the following steps.
First, we define a per run normalized activity $\widetilde{C}_{s,j}$:
\begin{equation}
\widetilde{C}_{s,j} =   \dfrac{C_{s,j}}{\frac{ \Delta t}{n T_{end} }\sum\limits_{s,j} C_{s,j}}
\end{equation}

Using the normalized activities we can define a new patternedness measure $P$, which detects how activity deviates from homogeneity during each time steps and then calculates the root mean square: 
\begin{equation}
P = \sqrt{\dfrac{\sum\limits_{j}\sum\limits_{s}(\widetilde{C}_{s,j} - \bar{C}_{j})^{2}}{(n \cdot (T_{end} / \Delta t) - 1)}},
\end{equation}
where $\bar{C}_{j}=\frac{1}{n}\sum\limits_{s}\widetilde{C}_{s,j}$

A high $P$ implies low homogeneity over time and ring-shaped segments and thus strong patternedness over time and segments, allowing us to easily compare these values for various experimental conditions.

This measure compares the activity of a given segment not to the overall mean activity level but to the activity level in a particular time step. This way we disqualify situations where the whole body shows high and homogeneous activity in one time interval and low homogeneous activity in other time intervals (`flickering' behaviour).

\bibliographystyle{apalike}

\clearpage
\subsection*{Supplementary Material 1: Mikado implementation details}
\label{sec:mikado}
Regarding implementation details, consider the case of cells arranged in a plane. Every elongated cell extends an axodendritic elongation $p$ with length $l_{p}$ in an arbitrary direction $\phi_{p}$, relative to the x-axis. Where elongations cross each other they create bidirectional connections between the originating cells. Given any two cells $i$ and $j$ with coordinates $(x_{i},y_{i})$ and $(x_{j},y_{j})$ and elongation directions relative to the x-axis $\phi_{i}$ and $\phi_{j}$ respectively, a triangle emerges, with at one point cell $i$, another point cell $j$ and the third point the crossing of their elongations. Two sides of this triangle consist of the elongations themselves. The third side, $d$, is the line between the cells, with an angle relative to the x-axis: $\phi_{d}$. Assuming $\phi_{i} \neq \phi_{j} \neq \phi_{d}$\footnote{In practice, given the granularity of the random numbers used, this assumption always holds.}, the sine-rule allows us to locate the crossing location, given the angles and the locations of the cells.
\begin{equation}
\frac{d}{\sin(\phi_{j}-\phi_{i})} = \frac{\rho_{i}}{\sin(\pi - \phi_{j}+\phi_{d})} = \frac{\rho_{j}}{\sin(\phi_{i}-\phi_{d})}
\end{equation}
Furthermore,
\begin{equation}
\sin(\pi-\phi_{j}+\phi_{d}) = -\sin(-\phi_{j}+\phi_{d}) = \sin(\phi_{j}-\phi_{d})
\end{equation}
Cell coordinates allow us to calculate $d$ and $\phi_{d}$:
\begin{equation}
d = \sqrt{(x_{i}-x_{j})^{2} + (y_{i}-y_{j})^{2}}
\end{equation}
\begin{equation}
\phi_{d} = \arctan  \Big( \frac{y_{j}-y_{i}}{x_{j}-x_{i}} \Big)
\end{equation}
Then we can solve for $\rho_{i}$ and $\rho_{j}$:
\begin{equation}
\rho_{i} = d \cdot \frac{\sin(\phi_{j}-\phi_{d})}{\sin(\phi_{j}-\phi_{i})} 
\end{equation}
\begin{equation}
\rho_{j} = d \cdot \frac{\sin(\phi_{i}-\phi_{d})}{\sin(\phi_{j}-\phi_{i})}
\end{equation}
If, for a given pair of cells, $\rho_{i}$ and $\rho_{j}$ are shorter than $l_{p}$, so $0 < \rho_{i} < l_{p}$ and $0 < \rho_{j} < l_{p}$ , the cells are connected. We loosened this constraint to  $-\tfrac{1}{2} < \rho_{i} < l_{p}$ and $-\tfrac{1}{2} < \rho_{j} < l_{p}$ to account for crossings occurring on the cell body of the originating cell.

All the above, however, is predicated on an infinite plane, while our system is finite and tube-shaped. Although a tube is essentially a curved 2-dimensional plane, its contiguousness results in some particular cases: elongations which may wrap around the cylinder. If we assume the cylinder is formed by making the x-direction periodic, we should consider not only the crossing of elongations starting at $(x_{i},y_{i})$ and $(x_{j},y_{j})$ but also at $(x_{i} + k_{i} \cdot \odot,y_{i})$ and $(x_{j} + k_{j} \cdot \odot,y_{j})$, with $k$ representing the number of windings and $\odot$ the circumference of the tube. We can limit the number of windings we need to consider, since the maximum elongation distance is limited: $|k| < (2 \cdot l_{p} / \odot) + 1$. The $+1$ is needed because the nearest copy might be on the other side of the tube's period. Another particular of the tube as opposed to a torus is that the crossing point may lie off the tube on the finite end, so we need to check whether the y-coordinate of the crossing, $y_{crossing} = \rho_{i} \cdot \sin \phi_{i} + y_{i}$ is on the tube: $-0.5 \leq y_{crossing} \leq (l_{tube}-1)\cdot \tfrac{1}{2} \sqrt{3} + 0.5$, where $l_{tube}$ is the number of rings on the tube and $\tfrac{1}{2} \sqrt{3}$ is the height of an equilateral triangle with sides of length 1.

\clearpage

\subsection*{Supplementary Material 2: Additional notes on pattern quantification}
\label{sec:segment_number}
The constant $n$, denoting the number of segments into which the modelled body is divided, requires more scrutiny. The analysis yields different outcomes for different values of $n$. This supplement aims to provide an explanation of why we chose to set  the number of segments to 16 instead of iterating over it or choosing another value. 

First of all, we use a power of two when setting the length of our model. We force our segments to all have the same length and thus  our possible choices for the number of segments are limited to powers of two as well. The shortest length used is $32$, that is $2^5$ and thus we have $2,4,8,16$ and $32$ to pick from. The values $2$ and $4$ are theoretically allowable but do not fit our intuition on wave propagation. The value $32$ is also allowed on theoretical grounds, but already nearest neighbor connections alone will spread out of the bin within a single time bin. This leaves us with two prima facie suitable candidates: $n=8$ and $n=16$, which we evaluated here.

\begin{figure}
\begin{center}
\includegraphics[width=.9\linewidth]{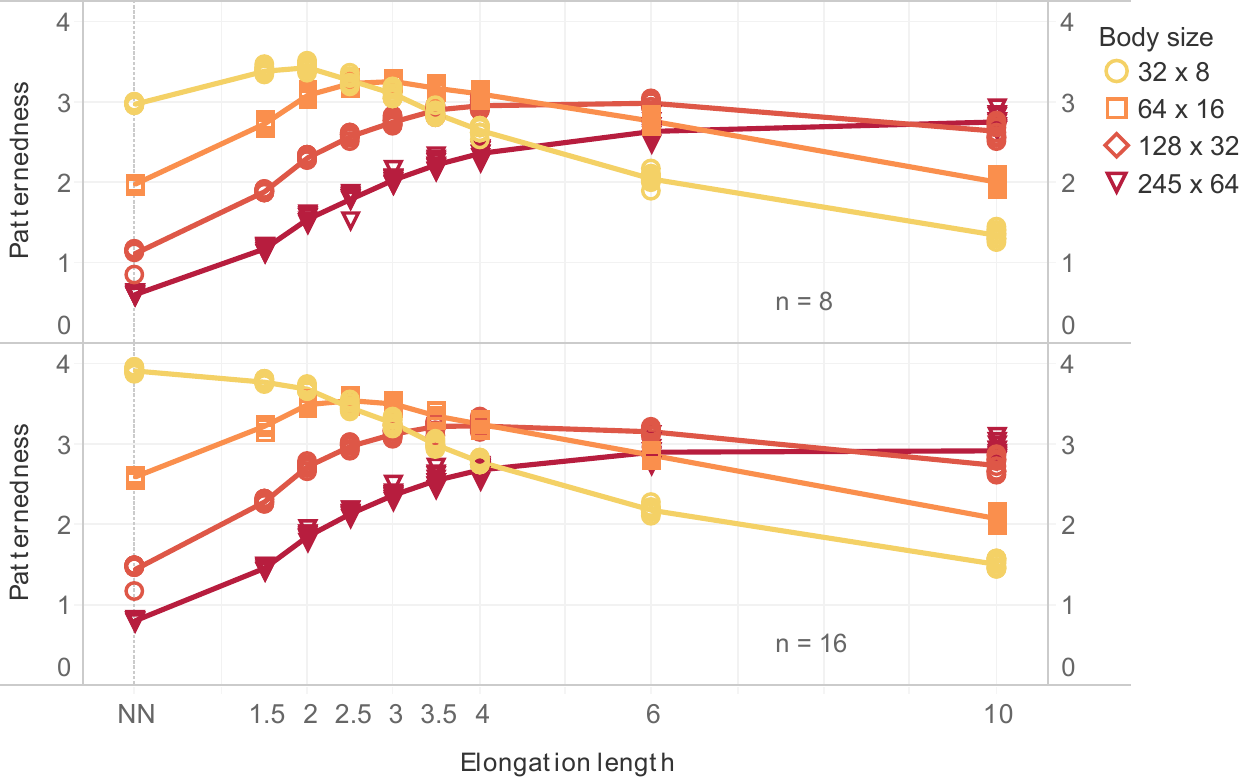}
\end{center}
\caption{Illustration of variation in segment number affecting the patternedness measure. Individual graphs represent segmentation options: the top graph represents 8 segments whereas the bottom graph represents 8 segments. Different body sizes are differentiated by color and point shape. Again, patternedness is shown on the y-axis, the different elongation lengths are on the x-axis. The points scattered around the graph indicate the 10 individual model iterations which were performed for each condition and the line represents the average.}
\label{fig:segmentation}
\end{figure}

Succinctly put, the qualitative results are similar when using different number of segments. Figure~\ref{fig:segmentation} shows that there is a difference in actual values but hardly a difference in overall shape. The effect of a lower number of segments is twofold: first, it causes patternedness to be higher for faster ring-shaped patterns. The optimum values will thus be biased towards longer elongation lengths, as those provide faster pattern propagation. Second, since more segments allows finer measurement the higher number of segments is able to show higher values overall. Even neat patterns within a large segment result in some lost signal. This also results in the effect visible in figure~\ref{fig:segmentation}, where 8 segments (top graph) for small systems (yellow circles) produces an improvement in patternedness between not having elongations at all and having elongations of length 2, since the broader patterns generated by elongations allow the segments to fill out whereas smaller segments get filled out even with patterns 1 cell wide.
\newpage
Since the aim of our study is to establish qualitative changes in whole-body coordination, having a measure that is constant across multiple body sizes is important, and dividing the body-tube in a fixed number of segments accomplishes this. Whether that is 8 or 16 is of secondary importance for the purposes of this study.

\end{document}